\pdfoutput=1
\documentclass[12pt]{article}
\usepackage[utf8]{inputenc}

\usepackage[shortcuts]{extdash}
\usepackage[nosort]{cite}
\usepackage[nottoc,numbib]{tocbibind}

\usepackage{csquotes}
\usepackage[english]{babel}
\usepackage{graphicx} 
\usepackage[left=1.5cm,right=1.5cm,top=1.5cm,bottom=1.5cm,includefoot,footskip=1.5cm]{geometry}
\usepackage{amsmath}  
\usepackage{cancel}
\usepackage{slashed}
\usepackage[mathscr]{eucal}
\usepackage{rotating,indentfirst,array,varioref}
\usepackage{appendix,marginnote,tikz,pgf,mathtools}
\usepackage{setspace}
\usepackage{misccorr}
\usepackage{siunitx}
\usepackage{amsthm}
\usepackage{amssymb}
\usepackage{wrapfig}
\addto\captionsrussian{
  \renewcommand{\contentsname}%
    {Contents}%
}
\usepackage{hyperref}
\hypersetup{
    colorlinks=true,
    linktoc=all,    
    linkcolor=blue,
		citecolor=blue
}
\usepackage{titlesec}
\titleformat{\chapter}
  {\normalfont\LARGE\bfseries}{\thechapter}{1em}{}
\titlespacing*{\chapter}{0pt}{3.5ex plus 1ex minus .2ex}{2.3ex plus .2ex}

\usepackage{authblk}

\newcommand{\pd}{\partial}

\newcommand{\br}[1]{{\overline{#1}}}

\def\<{\left(}
\def\>{\right)}

\begin{document}

\title{Five-point correlation numbers in minimal Liouville gravity and matrix models}

\author[1,2] {A. Artemev\thanks{artemev.aa@phystech.edu}}
\author[1,3] {A. Belavin\thanks{belavin@itp.ac.ru}}

\affil[1]{Landau Institute for Theoretical Physics, 142432, Chernogolovka, Russia}

\affil[2]{Skolkovo Institute of Science and Technology, 121205, Moscow, Russia}

\affil[3]{Kharkevich Institute for Information Transmission Problems,  127994,
Moscow, Russia }


\maketitle
\begin{abstract}
In this article, we will show how
to use Zamolodchikov's higher equations of motion in Liouville field theory to explicitly calculate $N$-point correlation numbers in minimal Liouville gravity for $N>4$. We find the explicit expression for the 5-point correlation numbers and compare it  with calculations in the one-matrix models.
\end{abstract}
\vskip 10pt
\section{Introduction}

There exist two different approaches to 2D quantum gravity. 
First of them is the "continuous" approach. In this approach
the theory is given by  the functional integral over 
2D Riemannian metric and the matter fields of 
a model of 2D conformal field theory (CFT) with central charge $c$.
After the conformal gauge fixing we get the theory which is called
 Liouville gravity \cite{P}. If the matter sector is described by one of the  minimal models of CFT \cite{BPZ}, then the theory is called minimal Liouville gravity (MLG) \cite{KPZ, DDK}.  

The second "discrete" approach, defined via the integral over $N\times N$ matrices, with $N$ tending to infinity, is called matrix models (MM) approach (see \cite{GM} and references there).

Both approaches is based  on the same idea of fluctuating geometry.
Therefore we can expect that they are identical.
Indeed, this conjecture has been confirmed  by a few explicit calculations 
in \cite {MSS}, \cite{KPZ}, \cite{GL},\cite{ABAZ}.
These tests were performed by comparing the
gravitational dimensions and correlation numbers in both 
(MLG and MM) approaches.
Unfortunately, these checks were only made for 2-, 3- and 4-point correlators
since only these cases were calculated on the MLG side\cite{AlZ}, 
\cite{ABAlZ}.
At the same time on the MM side the expressions for $N$-point correlators for any $N$ are known in one-matrix model case \cite{ABAZ}.

Therefore in this work we consider the problem of how to compute the 
$N$-point correlation numbers in  MLG when $N>4$
(we will focus on the case when the 2D worldsheet
is topologically a sphere.)

$N$-point correlation numbers in MLG are defined as  vacuum averages of the product of physical (i.e. BRST-closed) observables.
Physical observables of MLG  appear either as local  BRST-closed fields located  at the point $x$ on the sphere, or as integrals of local densities 
($(1,1)$- forms) over their positions.

We will consider  two kinds of local BRST-closed fields in MLG.
We denote them by $W_{m,n}(x)$ and $O_{m,n}(x)$.
The local densities whose integrals are  physical observables will be 
denoted by $U_{m,n}(x)$.
We remind the definition of these objects and  the relations between them
in the next section.
Below we will consider the correlators which only include the product of  
$W_{m,n}(x)$ fields and  integrals of the $U_{m,n}(x)$ densities. 

The fields $W_{m,n}(x)$ have ghost numbers equal to one, and the ghost numbers of fields $U_{m,n}(x)$ are equal to zero. The correlators of the considered fields on the sphere do not vanish only if the total ghost number is $3$. 

It follows that the $N$-point correlator must contain three fields
of the type $W_{m_i,n_i}(x_i), i=1,2,3$ and $(N-3)$ integrated fields of type
$U_{m_i,n_i}(x_i), i=4,...,N$.
The 3-point correlation numbers do not contain the fields  $U_{m_i,n_i}(x )$ at all. To find them, one only needs to know the three point functions 
in minimal models of CFT\cite{BPZ} and in Liouville field theory
\cite{DO}, \cite{ZZ}.
The four-point correlator contains one integration over the position of 
$U_{m,n}(x)$.
In \cite{ABAlZ} a way for computing the moduli integrals was
developed by Alexei Zamolodchikov and one of the authors.
Using the higher Liouville equations of motion (HEM) \cite {AlZ} allows one to reduce the moduli integrals to the boundary terms. 
This approach was applied for the four-point correlation numbers in
 minimal Liouville gravity and minimal super-Liouville gravity in \cite{ABAlZ},\cite{ABVB},\cite{KAVB}.
In this case, the fact was used that, besides the field $U_{m,n}(x)$, the other three fields in the correlator $W_{m_i,n_i}(x_i)$ are BRST-closed.
It follows that the BRST exact terms in the r.h.s. of  the HEM relation for the field $U_{m,n}(x)$ can be neglected in this case.
The integral of the remaining term is reduced to computable boundary contributions from the vicinity of points $x_i, i=1,2,3$ 
and $\infty$.

The situation in the case when $N>4$ is different, since, in addition to three BRST closed fields $W_{m_i,n_i}(x_i)$, the vacuum average includes $N-3>1$ integrals of  $U_{m,n}(x)$. In this case, BRST-exact terms on the right side of the HEM relation for one of two (or more)  $U_{m,n}(x)$  fields can not be neglected.

However, as we will demonstrate below, these contributions, added to the contribution of the main (not $\mathcal{Q}$-exact) term, reduce the entire expression to a sum of boundary contributions, which have the same form as
in a $4$-point correlator.

It is important that the boundary contributions from the vicinity of other integrated operator insertions $U_{m,n}(x)$ have also the form similar to the contribution of local fields $W_{m_i,n_i}(x_i)$.


The results of MLG were first tested   against the corresponding correlation numbers from the matrix models   by Moore, Seiberg, and Staudacher (MSS) \cite{MSS}.
In particular, considering the case of the
$p$-critical point of the one-matrix model and the corresponding $(2,2p+1)$ minimal Liouville gravity, MSS partially determined the
resonance terms in the relation between the coupling parameters, which allowed them to establish equivalence up to the level of two-point correlation numbers.

In \cite{ABAZ} the agreement between the Matrix Models and minimal Liouville gravity results has been reached up to the level of four-point correlation numbers by demanding that the higher order correlation numbers satisfy the fusion rules inherent to the MLG.
 In the process, the higher order resonance terms were determined
from this requirement. 
In \cite{ABAZ} the full resonance transformation
which relates coupling parameters in the $p$-critical  one-matrix models and $(2,2p+1)$ minimal Liouville gravity was conjectured.
But this conjecture has been checked against MM only up to 4-point numbers since the results for higher correlation numbers in MLG were not available at that time.
G. Tarnopolsky \cite{GT} continued investigations of $(2,2p+1)$ minimal gravity using the conjecture of \cite{ABAZ} and obtained the explicit expression for the five-point correlation numbers in one-matrix model.
He checked that the correlation numbers satisfy the necessary fusion rules. Since in this work we obtained the explicit expression for 5-point case in  MLG, we can compare it to the results of \cite{GT}.

The article is organized in the following way. 
In section 2, we recall the known facts about MLG that we need for further calculations. In section 3, we present an approach to computing N-point correlation numbers in MLG with $N>4$. Then we explicitly compute a 5-point correlation number in MLG. In section 4, we compare the  expression for the 5-point correlation numbers in MLG and compare it with the  expressions for the 5-point correlation  numbers in MM given in \cite{GT}.


\section{Preliminaries}

The minimal Liouville gravity (MLG) is a special case of the  Liouville gravity\cite{P}. 
This is a  CFT of total central charge equal to $0$  consisting of  Liouville field theory (LFT) describing  the gravity sector, a  minimal model $\mathcal{M}_{q',q}$ of CFT \cite{BPZ} for the matter sector, and the reparametrization BRST ghosts $B,C$ CFT of central charge $-26$:
\begin{equation}
    A_{MLG} = A_L + A_{\mathcal{M}_{q',q}} +\underbrace{\frac{1}{\pi}\int d^2x\,\left( C \br{\pd} B + \br{C} \pd \br{B} \right)}_{A_{ghost}} \label{mlglag}.
\end{equation}

 The central charge of Liouville theory is defined by $(q',q)$ --- characteristics of the minimal model. It follows from the requirement of vanishing total central charge of the theory.

\vskip 5pt

{\bf  $\mathcal{M}_{q',q}$ matter sector.}

Minimal models of CFT $\mathcal{M}_{q',q}$
\cite{BPZ} are consistently defined if the parameters  $q$ and $q'$ are coprime integers.
 In this case the finite set of Virasoro irreducible representations consisting of degenerate primary fields $\Phi_{m,n}$ with $1\leq m<q $ and $1\leq n<q'$ and their descendants form the whole space of states of $\mathcal{M}_{q',q}$ model. It is self-consistent, i.e. satisfies all axioms of the conformal bootstrap, and is an exactly solvable CFT. 
In what follows, we will consider  only the  models of such type.
Let us denote by $b^{2}$ the parameter $q'/q$. Then $\mathcal{M}_{q',q}$ has central charge
\begin{equation}
c=1-6(b^{-1}-b)^{2} \label{cM}%
\end{equation}
and the degenerate primary fields $\Phi_{m,n}$ have dimension 
\begin{equation}
  \Delta^M_{m,n}= -(b^{-1} - b)^2/4 + \lambda^2_{m,-n}
\end{equation}
where yet another convenient notation
\begin{equation}
\lambda_{m,n}=(mb^{-1}+nb)/2 \label{lmn}%
\end{equation}
is introduced. We will also use notation $\Phi_\alpha$ to denote minimal model primary fields of dimension $\Delta^{(M)}_\alpha = \alpha(\alpha-b^{-1}+b)$.
The main restrictions, which finally fix the  construction of the minimal model are as follows:
 
 \begin{enumerate}
     \item The degenerate fields $\Phi_{1,2}$ and $\Phi_{2,1}$ (and therefore in general the whole set $\left\{  \Phi_{m,n}\right\}$) are in the spectrum;
     \item The null-vector in the degenerate representation $\Phi_{m,n}$ vanishes for all $m,n$
\begin{equation}
D_{m,n}^{\text{(M)}}\Phi_{m,n}=\bar D_{m,n}^{\text{(M)}}\Phi_{m,n}=0.
\label{GMM}%
\end{equation}
Here $D_{m,n}^{\text{(M)}}$ ($\bar D_{m,n}^{\text{(M)}}$) are the operators
made of the holomorphic (antiholomorphic) Virasoro generators $L^M_{n}$ ($\bar L^M_{n}$),
which create the singular vector on level $mn$ in the Virasoro module of $\Phi_{m,n}$. 
     \item The identification $\Phi_{q-m, q'-n} = \Phi_{m, n}$ is also  assumed.
 \end{enumerate}
It turns out that these  definitions impose severe restrictions on
the structure of the theory.
In particular, the three-point function of primary fields
can be unambiguously recovered from these requirements.
\vskip 10pt

{\bf Liouville field theory.}

LFT is  the quantized version of the classical theory based on the Liouville action. LFT is  a conformal field theory with central charge $c_{\text{L}}$. We 
parametrize it in terms of variable $b$ or
\begin{equation}
Q=b^{-1}+b \label{Q}%
\end{equation}
as
\begin{equation}
c_{\text{L}}=1+6Q^{2} \label{cL}%
\end{equation}
In MLG from the requirement of vanishing total central charge it follows that $b$ is the same as the parameter of the minimal model defined in the previous subsection, which is why we denote it by the same letter.

The parameter $b$ enters the local Lagrangian
\begin{equation}
\mathcal{L}_{\text{L}}=\frac1{4\pi}\left(  \partial_{a}\phi\right)  ^{2}+\mu
e^{2b\phi} \label{LFT}%
\end{equation}
where $\mu$ is the scale parameter called the cosmological constant and $\phi$
is the dynamical variable for the quantized metric
\begin{equation}
ds^{2}=\exp\left(  2b\phi\right)  \widehat{g}_{ab}dx^{a}dx^{b} \label{ds}.%
\end{equation}
Here $\widehat{g}_{ab}$ is the "background"  metric. 
 Basic primary fields are the exponential operators $V_{a}\equiv \exp\left(  2a\phi\right)
$, parameterized by a continuous (in general complex) parameter $a$ in the way
that the corresponding conformal dimension is
\begin{equation}
\Delta_{a}^{\text{(L)}}=a(Q-a) \label{DL}%
\end{equation}

In what follows, two types of primary fields of Liouville sector will play an important role in constructing the physical fields of the MLG.

The first kind are degenerate  primary fields $V_{m,n} \equiv V_{a_{m,n}}$ with
\begin{equation}
a_{m,n} = - b^{-1} \frac{(m-1)}{2} - b \frac{(n-1)}{2} \label{lioprim}.
\end{equation}
These fields satisfy  equations 
$D^{(L)}_{m,n} V_{m,n}=\bar D^{(L)}_{m,n}V_{m,n}=0$ analogous to the ones in minimal models.

The second kind are the primary fields $V_{m,-n}$, whose role together with the ghost field $C$  is to dress the  primary fields $\Phi_{m,n}$ of the matter sector and get as a  result a BRST-closed field $W_{m,n}$. We will show it below.

Liouville field theory is exactly solvable \cite{DO},\cite{ZZ}. 
The three-point correlation function 
$C_{\text{L}}(a_{1},a_{2},a_{3})=\left\langle V_{a_{1}%
}(x_{1})V_{a_{2}}(x_{2})V_{a_{3}}(x_{3})\right\rangle _{\text{L}}$ is known
explicitly for arbitrary exponential fields
\begin{equation}
C_{\text{L}}(a_{1},a_{2},a_{3})=\left(  \pi\mu\gamma(b^{2})b^{2-2b^{2}%
}\right)  ^{(Q-a)/b}\frac{\Upsilon_{b}(b)}{\Upsilon_{b}(a-Q)}\prod_{i=1}%
^{3}\frac{\Upsilon_{b}(2a_{i})}{\Upsilon_{b}(a-a_{i})} \label{CL}%
\end{equation}
where $a=a_{1}+a_{2}+a_{3}$ and $\Upsilon_{b}(x)$ is a special function
related to the Barnes double gamma function \cite{ZZ}.

The local structure of LFT is  completely determined by 
the general ``continuous'' operator product expansion (OPE) of  Liouville exponential fields
\begin{equation}
V_{a_{1}}(x)V_{a_{2}}(0)=\int^{\prime}\frac{dP}{4\pi}C_{a_{1},a_{2}%
}^{\text{(L)}Q/2+iP}(x\bar x)^{\Delta_{Q/2+iP}^{\text{(L)}}-\Delta_{a_{1}%
}^{\text{(L)}}-\Delta_{a_{2}}^{\text{(L)}}}\left[  V_{Q/2+iP}(0)\right]
\label{LOPE}%
\end{equation}
where the structure constant is expressed through 
(\ref{CL}) 
$C_{a_{1},a_{2}}^{\text{(L)}p}=
C_{\text{L}}(g,a,Q-p).$ 
The integration contour here is the
real axis if $a_{1}$ and $a_{2}$ are in the ``basic domain''
\begin{equation}
\left|  Q/2-\operatorname*{Re}a_{1}\right|  +\left|  Q/2-\operatorname*{Re}%
a_{2}\right|  <Q/2 \label{basic}%
\end{equation}

In other domains of these parameters an analytic continuation is implied, which means that the integration contour should be deformed or, equivalently, we should add to (\ref{LOPE}) separately residues in the poles which cross the real line during the analytic continuation from basic domain. These contributions in the OPE are referred to as "discrete terms" in the future; they are particularly important when one of the fields in the OPE is degenerate.
\vskip 5pt

{\bf  Ghost field theory. BRST invariance.}

The ghost sector is the  fermionic $BC$ system of spin $(2,-1)$%
\begin{equation}
A_{\text{gh}}=\frac1\pi\int(C\bar\partial B+\bar C\partial\bar B)d^{2}x
\label{Agh}%
\end{equation}
with central charge $-26$, which corresponds to the gauge fixing Faddeev-Popov
determinant. 
The matter+Liouville stress tensor $T$ is a generator of $c=26$ Virasoro algebra. 
Together with the ghost field theory this forms a BRST complex with respect to the nilpotent BRST charge, the holomorphic part of which is
\begin{equation}
\mathcal{Q}=\oint\left(  CT+C\partial CB\right)  \frac{dz}{2\pi i}.
\label{brst}%
\end{equation}
 By definition the physical fields of MLG belong to BRST cohomogy of the 
charge $\mathcal{Q}$ and its antiholomorphic part $\mathcal{\bar{Q}}$.
\vskip 5pt

{\bf  Physical (BRST-closed) fields and  their correlators.}

The simplest cohomology representatives of ghost number zero can be obtained by dressing minimal model primaries $\Phi_{m,n}$ with Liouville fields 
$V_{m,-n}$ so that their total conformal dimension is (1,1) and then integrating the obtained fields $U_{m,n} \equiv V_{m,-n}\Phi_{m,n}$ over the surface. 
The variation of $U_{m,n}$ is a full derivative
\begin{equation}
\mathcal{Q} U_{m,n} = \pd (C U_{m,n}) \label{ubrstvar}
\end{equation}
so such  fields integrated over the sphere are BRST invariant 
(subtleties connected with boundary terms could emerge depending on the other insertions). 

To get physical states of ghost number $1$, instead of integrating, one can dress the $U_{m,n}$ field with the  ghost fields $C$, $\br{C}$  and obtain the $(0,0)$ form $W_ {m,n} \equiv C \br{C} U_{m,n}$ which is BRST-closed, 
$\mathcal{Q} W_{m,n}=\mathcal{\br{Q}} W_{m,n} = 0$.  
We will also in the future denote these fields by their Liouville parameter 
$a$ : $W_a = V_a \Phi_{a-b}$.

 If we are interested in correlators of multiple operators 
$\int d^2x\,U_{m,n}(x)$ and $W_{m,n}(x)$ on a sphere, the ghost number anomaly (presence of $C$-zero modes of kinetic operator in ghost action) requires number of $C$-ghosts in such correlator to be equal to three. Thus we need to insert three $W_{m_i,n_i}(x_i), i=1,2,3$ fields at some points $x_1, x_2, x_3$ and all the other operators should be integrals of densities $U_{m,n}(x)$.

In minimal Liouville  gravity, there is an additional  set  of  BRST-closed fields with ghost number zero that form  the so-called "ground ring"  
\cite{W}. 
These fields have the general form
\begin{equation}
O_{m,n}(x) = H_{m,n} \br{H}_{m,n} \Theta_{m,n}, \, \   \ 
\Theta_{m,n} \equiv V_{m,n} \Phi_{m,n}. 
\end{equation}
Here $H_{m,n}$ is a polynomial of degree $mn-1$ of Virasoro generators and ghosts $B$ and $C$. 
The general form for $H_{m,n}$ is unknown, but it can be found case by case by requiring $\mathcal{Q}$-closedness of the operator $O_{m,n}$. 

The polynomials $H_{m,n}$  play an important  role in the derivation of the so-called higher equations of motion (HEM) of Al. Zamolodchikov and the key properties of the physical fields $W_{m,n}(x)$ and $U_{m,n}(x)$.

We quote the expressions for $H_{m,n}$ for the first couple of cases 
\begin{align}
&H_{1,2} = L^M_{-1} - L_{-1} + b^2 CB \label{h12}    \\
&H_{1,3} = (L^M_{-1})^2 -L^M_{-1} L_{-1} + L_{-1}^2 - 2 b^2 (L^M_{-2} - L_{-2}) + 2 b^2 CB (L^M_{-1} - L_{-1}) -4b^4 C \pd B  \label{h13}
\end{align}
where by $L_n^M$ are denoted Virasoro generators of the matter Minimal model and by $L_n$ of the Liouville theory.
\vskip 5pt

Some properties of the ground ring operators include:
\begin{enumerate}
    \item independence of the correlator on their position in the sense that
\begin{equation}
\pd O_{m,n} = \text{BRST-exact}.
\end{equation}
This is valid for any BRST-closed operator since we have
\begin{equation}
    \pd = L_{-1}^{L+M} + L_{-1}^{gh} = \lbrace \mathcal{Q}, B_{-1} \rbrace
\end{equation}

\item Fusion of two operators $O$ is very simple in cohomology:
\begin{equation}
O_{m,n}(x) O_{m',n'}(0) = \sum \limits_{r = |m-m'|+1:2}^{m+m'+1} \sum \limits_{s = |n-n'|+1:2}^{n+n'+1} G_{r,s}^{(m,n)|(m',n')} O_{r,s}(0) + 
\text{BRST-exact}.\label{ooope}
\end{equation}

\item Similarly for the fusion with ghost number 1 operators $W_a$ we have
\begin{equation}
O_{m,n} W_a = \sum \limits_{r = -m+1:2}^{m-1} \sum \limits_{s = -n+1:2}^{n-1} A_{r,s}^{(m,n)} (a) W_{a + \frac{rb^{-1} + s b}{2}} +\text{BRST-exact} \label{owope}
\end{equation}
\end{enumerate}

The algebra of the ground ring operators is such that both coefficients
$G_{r,s}^{(m,n)|(m',n')} $ and $A_{r,s}^{(m,n)} (a) $ can be put to one  with renormalization of the operators $O_{m,n}$ and $W_{m,n}$.

In fact, the following formulas are valid
\begin{equation}
 G_{r,s}^{(m,n)|(m',n')} = \frac{\Lambda_{m,n} \Lambda_{m',n'}}{\Lambda_{r,s}};\,  \Lambda_{m,n} = \frac{B_{m,n}}{\pi} \mathcal{N}(a_{m,-n})
\end{equation}
\begin{equation}
    A_{r,s}^{(m,n)}(a) = \frac{B_{m,n}}{\pi} \frac{\mathcal{N}(a) \mathcal{N}(a_{m,-n})}{\mathcal{N}(a+\lambda_{r,s})} \label{anormf}
\end{equation}
with $B_{m,n}$  defined in the following subsection (\ref{bmn}) 
and

\begin{equation}
\mathcal{N}(a) =\frac{\pi}{(\pi\mu)^{a/b}}
\left[\frac{\gamma(2ab-b^2)\gamma(2ab^{-1}-b^{-2})} 
{\gamma^{2a/b-1}(b^{2})\gamma(2-b^{-2})}\right]^{1/2}.
\end{equation}


So, after renormalizing $\mathcal{O}_{m,n} = \Lambda^{-1}_{m,n}O_{m,n}$
and \, $\mathcal{W}_a = {\mathcal{N}(a)}^{-1} W_a$, 
both $G$ and $A$ become equal to one and we get (up to BRST-exact terms)
\begin{equation}
\mathcal{O}_{m,n} \mathcal{O}_{m',n'} = 
\sum \limits_{r = |m-m'|+1:2}^{m+m'+1} 
\sum \limits_{s = |n-n'|+1:2}^{n+n'+1} \mathcal{O}_{r,s} .\label{ooope}
\end{equation}

Similarly for the fusion $\mathcal{O}_{m,n}$ with the  operator $\mathcal{W}_a$ 
we have
\begin{equation}
\mathcal{O}_{m,n} \mathcal{W}_a = \sum \limits_{r = -m+1:2}^{m-1} 
\sum \limits_{s = -n+1:2}^{n-1}  \mathcal{W}_{a + \frac{rb^{-1} + s b}{2}}. \label{owope}
\end{equation}

\vskip 10pt

{\bf  HEM and key relations of MLG.}

The important progress in computation of the  integrals over the moduli space 
 was achieved using higher equations of motion (HEM) of  Al. Zamolodchikov\cite {HEM}. HEM are a set of operator relations in Liouville field theory.
They involve the so-called logarithmic fields $V'_a $ and $O'_{m,n}$, 
where
\begin{equation}
V'_a(x) = \frac{1}{2} \frac{\pd}{\pd a} V_a(x).
\end{equation}
These fields are called logarithmic since its OPE with ordinary primary operators generally involve logarithms.
We will denote by $V'_{m,n}$ such logarithmic operator evaluated at the point corresponding to degenerate dimension $a=a_{m,n}$. 

The HEM equate an descendant of logarithmic operators to some multiple 
of $V_{m,-n}$. 
They look as follows \cite{HEM}
\begin{equation}
  D^{(L)}_{m,n}  \br{D}^{(L)}_{m,n}  V'_{m,n} = B_{m,n} V_{m,-n}
\end{equation}
\begin{equation}
    B_{m,n} = (\pi \mu \gamma(b^2) b^{2-2b^2})^n \frac{\Upsilon'_b(2\alpha_{m,n})}{\Upsilon_b(2\alpha_{m,-n})} \label{bmn}
\end{equation}
\vskip 5pt
As shown in \cite {ABVB} for super-LG case using HEM (but it is also valid in the case without supersymmetry) there exists the following {\bf key relation} between the BRST-closed field 
$W_{m,n}$
and  the field $O'_{m,n}$ which is a  "logarithmic counterpart" of the element  $O_{m,n}$ 
\begin{equation}
W_{m,n}= B^{-1}_{m,n} \br{\mathcal{Q}}\mathcal{Q} O'_{m,n} \label{keyrel1}
\end{equation}
where
$
O'_{m,n}:= H_{m,n} \br{H}_{m,n}\Theta'_{m,n}$ and $\Theta'_{m,n}:=\Phi_{m,n}V'_{m,n}$.
\vskip 5pt 

The relation (\ref{keyrel1}) seems "strange", since on its l.h.s. we see a non-trivial BRST-closed  element, but the element on r.h.s. looks  BRST-exact.
However, there is no contradiction here, because the logarithmic fields
$V'_{m,n}$ and $O'_{m,n}$ do not belong to the space where the BRST operator
$\mathcal{Q}$ is  defined. Here it acts on some extension of this space.

Using  the relations $W_{m,n}=\bar{C}C U_{m,n}$,
  $L^M_{-1}+L_{-1}= B_{-1}\mathcal{Q}+\mathcal{Q}B_{-1}$ and $B_{-1}C(z)=I$,
	(here 	$B_{-1}$ is a Fourier component  of the ghost $B(z)$),  
	we can derive from (\ref{keyrel1}) the second {\bf key relation} (see also \cite {ABAlZ})

	\begin{equation}
	U_{m,n}= B^{-1}_{m,n}(\br{\pd} - \bar{\mathcal{Q}}\bar{B}_{-1}) 
	    (\pd  - \mathcal{Q}B_{-1})O'_{m,n}. \label{keyrel2}
	\end{equation}
	
At last, using the relation (\ref{keyrel2}) we can perform the explicit calculations of 	$N$-point correlation functions in MLG, reducing step by step 	$(N-3)$ integrals of type $\int {U_{m,n}(x)}d^2x$ in the correlators to the boundary contributions in the  positions of the other fields and in $\infty$.

\vskip 10 pt

{\bf Calculation of four-point correlation number.}
We will calculate the correlator
\begin{equation}
\frac{1}{Z_L} \left\langle \int d^2x\,U_{m,n}(x) W_{a_1}(x_1) W_{a_2}(x_2) W_{a_3}(x_3) \right\rangle
\end{equation}
This was first done in \cite{ABAlZ}. 
Such a correlator will be considered thinking of matter fields $\Phi_{a_i-b}$ as fields with generic conformal dimension (i.e. OPEs containing them and degenerate operators have maximally possible number of terms); only for the integrated operator matter dimension is significantly degenerate. This allows to not care about subtleties in OPE when we will need to use it. 
We will also separate the normalization factors $\prod \limits_{i=1}^3 \mathcal{N}(a_i) \times \mathcal{N}(a_{m,-n})$ and, in fact, consider the correlator of normalized operators
\begin{equation}
C_4(a_1,a_2,a_3 |m,n) \equiv \frac{1}{Z_L} \left\langle \int d^2x\,\frac{U_{m,n}(x)}{\mathcal{N}(a_{m,-n})} \mathcal{W}_{a_1}(x_1) \mathcal{W}_{a_2}(x_2) \mathcal{W}_{a_3}(x_3) \right\rangle
\end{equation}

First, we will rewrite the integral operator $U_{m,n}$ according to the equation (\ref{keyrel2}). Since $W$-operators are $\mathcal{Q}$-closed, we ignore BRST-exact terms that appear there and write
\begin{align}
&Z_L\,C_4(a_1,a_2,a_3 |m,n) = \frac{1}{B_{m,n} \mathcal{N}(a_{m,-n}) }\left\langle \int d^2x\,\pd \br{\pd} O'_{m,n}\, \mathcal{W}_{a_1}(x_1) \mathcal{W}_{a_2}(x_2) \mathcal{W}_{a_3}(x_3)  \right\rangle = \nonumber \\
& =\frac{1}{\pi}\left\langle \int d^2x\,\pd \br{\pd}\mathcal{O} '_{m,n}\, \mathcal{W}_{a_1}(x_1) \mathcal{W}_{a_2}(x_2) \mathcal{W}_{a_3}(x_3)  \right\rangle
\end{align}
(we switched to normalized operator $\mathcal{O}'_{m,n} = \frac{O'_{m,n}}{\Lambda_{m,n}}$). Now we have integral of full derivative in the correlator, and therefore we can use the Stokes formula to take it. It reduces to boundary contributions from the vicinity of points $x_i$ where other operators are inserted and infinity. These contributions are non-zero because of special nature of operator $O'_{m,n}$. Indeed,  in the OPE with $W$ there will be logarithmic terms, which after differentiation give delta-functions:
\begin{equation}
\pd \br{\pd} \log  (x\br{x}) = \pi \delta^{(2)}(x)
\end{equation}
(or, otherwise, after reducing integral to the contour one, logarithm gives $1/z$ after differentiation). We need thus to calculate these logarithmic contributions. 

First we deal with what comes from infinity. As follows from the Liouville OPE with stress-energy tensor, operator $V'_{m,n}$ for $x \to \infty$ behaves like (see \cite{ABAlZ})
\begin{equation}
V'_{1,2}(x) \sim - \Delta'_{m,n} \log (x \br{x}) V_{1,2}(0),\,\Delta'_{m,n} \equiv 2 \lambda_{m,n} =  m b^{-1} + n b
\end{equation}
Operator $O'_{m,n}$ behaves similarly; at infinity we can replace it with $O_{m,n}$ with this coefficient and the logarithm. Then, corresponding boundary contribution is
\begin{equation}
-2 \lambda_{m,n} \left\langle \mathcal{O}_{m,n}(0) \mathcal{W}_{a_1}(x_1) \mathcal{W}_{a_2}(x_2) \mathcal{W}_{a_3}(x_3)  \right\rangle
\end{equation}
We note that this expression does not depend on the position of $O_{m,n}$ since this field is BRST-closed, as are other fields in this correlator. For this reason we can perform OPE of $O$ with any of $W$ operators to obtain e.g.
\begin{equation}
-2 \lambda_{m,n} \sum \limits_{r=-m+1:2}^{m-1} \sum \limits_{s=-n+1:2}^{n-1} \left\langle  \mathcal{W}_{a_1 + \lambda_{r,s}}(x_1) \mathcal{W}_{a_2}(x_2) \mathcal{W}_{a_3}(x_3) \right\rangle
\end{equation}

To examine the logarithmic factors in the OPE of $O'$ with $W$, looking at Liouville OPE, we can note that the logarithmic factors can appear only when differentiating by $a$ power factors $(x\br{x})$ in the discrete terms. In every other factor we can put $a = a_{m,n}$ without trouble. For example, when considering such terms for $V_{1,2}'$ we obtain
\begin{equation}
 \log (x\br{x}) \left(q_{0,1}^{(1,2)}(a) (x \br{x})^{ab} C_L^+(a) [V_{a-b/2}(0)] + q_{0,-1}^{(1,2)}(a) (x\br{x})^{1-ab + b^2)}  C_L^-(a) [V_{a+b/2}(0)] \right)  \label{vprime12ope}
\end{equation}
\begin{equation}
q_{r,s}^{(m,n)} \equiv |a - \lambda_{r,s} - \frac{Q}{2}| - \lambda_{m,n}
\end{equation}
In other words, logarithmic part is similar to what we get for OPE with usual primary field $V_{1,2}$, but the terms are decorated by additional factors $q_{r,s}^{(1,2)}$. This is valid for arbitrary $V'_{m,n}$ as well. 
Multiplying Liouville OPE with OPE for minimal model and acting with operators $H_{m,n}$, in logarithmic terms only contributions from $W_{a - \lambda_{r,s}}$ will remain and coefficients $A_{r,s}^{(m,n)}$ will appear additionally. 
Thus, we obtain the following OPE:
\begin{equation}
\mathcal{O}'_{m,n}(x) \mathcal{W}_a(0) = \log (x \br{x}) \sum \limits_{r=-m+1:2}^{m-1} \sum \limits_{s=-n+1:2}^{n-1} q_{r,s}^{(m,n)}(a) \mathcal{W}_{a-\lambda_{r,s}} + \text{less singular terms} \label{OpWOPE}
\end{equation}
and contributions to the correlator
\begin{equation}
-\sum \limits_{i=1}^3\sum \limits_{r=-m+1:2}^{m-1} \sum \limits_{s=-n+1:2}^{n-1} q_{r,s}^{(m,n)}(a_i) \langle \mathcal{W}_{a_i - \lambda_{r,s}} \dots \rangle
\end{equation}
The additional minus sign appears because boundary component contours surrounding $\infty$ and points $x_i$ have opposite orientation. Now, using that all normalized three-point functions of $W$ operators become the same constant $-b^{-2}(b^{-4}-1)$ (\cite{AlZ},\cite{ABAlZ}), we arrive at the expression for total correlation function (normalized by partition function $Z_L$)
\begin{equation}
C_4(a_1, a_2, a_3| m,n) = -(b^{-6}-b^{-2}) \left[- 2 mn \lambda_{mn} - \sum \limits_{i=1}^3 \sum \limits_{r=-m+1:2}^{m-1} \sum \limits_{s=-n+1:2}^{n-1} q_{r,s}^{(m,n)}(a_i) \right].
\end{equation}

\section{Five-point correlator in MLG.}
 
Here we begin to extend the results obtained in MLG earlier to the case of higher multipoint correlators and consider 5-point correlation numbers in  $(2, 2p+1)$   MLG . We will assume that only two integrated fields are degenerate and are of the form $U_k \equiv U_{1,k+1}$:
\begin{equation}
C_5(a_1,a_2,a_3|k_1,k_2) = Z_L^{-1} \left\langle \int d^2x\, 
\frac{U_{1,k_1+1}(x)}{\mathcal{N}(a_{1,-1-k_1)}} \int d^2y\, \frac{U_{1,k_2+1}(y)}{\mathcal{N}(a_{1,-1-k_2)}}\,\mathcal{W}_{a_1}(x_1) \mathcal{W}_{a_2}(x_2) \mathcal{W}_{a_3}(x_3) \right\rangle.
\end{equation}
 As before, we assume that matter fields $\Phi_{a_i-b}$ in the three non-integrated operators have "generic" dimension.

We start the calculation by integrating over the variable $x$ and
using HEM  for the field $U_{k_1}$. 
The term with the full derivative $\pd \br{\pd} \mathcal{O}'_{m,n}$ can be reduced to boundary terms in the vicinity of $x_i$, $y$ and $\infty$.
\vskip 5pt
 
{\bf Contributions of  non-integrated fields $W_a(x_i)$.}

For $x_i$ contributions, we perform OPE of $\mathcal{O}'$ 
with $\mathcal{W}_a(x_i)$. As before, only the logarithmic terms are important. In total, these contributions are
\begin{equation}
- \sum \limits_{i=1}^3 \sum \limits_{s=-k_1:2}^{k_1} q_{0,s}^{(1,k_1+1)}(a_i) \left\langle \int d^2y\, \frac{U_{1,k_2+1}(y)}{\mathcal{N}(a_{1,-1-k_2)}}\,\mathcal{W}_{a_i-\lambda_{0,s}}(x_i) \dots \right\rangle \label{firstcont}.
\end{equation}
Therefore, the $x_i$ boundary contributions are expressed in terms of 4-point correlators with 3 generic and 1 degenerate fields that were calculated earlier.
\vskip 5pt 
{\bf Contribution from $x=\infty$  .}

 Contribution from $x=\infty$ in the integral over $x$ are given by
the following expression
\begin{equation}
    -2 \lambda_{1,k_1+1}  \int d^2y\ \left\langle \mathcal{O}_{1,k_1+1}(0)\, \frac{U_{1,k_2+1}(y)}{\mathcal{N}(a_{1,-1-k_2)}}\, \mathcal{W}_{a_1}(x_1) \mathcal{W}_{a_2}(x_2)  \mathcal{W}_{a_3}(x_3) \right\rangle \label{infcont}.
\end{equation}

To compute it we rewrite the second integrated operator $U_{1,k_2+1}(y)$ via HEM and integrate it by parts. The $\mathcal{Q}$-exact terms are irrelevant, since all other insertions are $\mathcal{Q}$-closed. 
After this we get 
\begin{equation}
-\frac{2 \lambda_{1,k_1+1}}{\pi} \int d^2y\,\pd \br{\pd}
\left\langle\left(\mathcal{O}'_{1,1+k_2}(y) \right) 
\mathcal{O}_{1,k_1+1}(0) \mathcal{W}_{a_1}(x_1) \mathcal{W}_{a_2}(x_2) \mathcal{W}_{a_3}(x_3) \right\rangle \label{continf},
\end{equation}
which is, as before, reduced to a sum of boundary terms in the vicinity of $0$, $x_i$ and contribution from infinity. 

The new thing we need to take into account is  a contribution of  OPE 
$O'_{1,1+k_2}(y) O_{1,k_1+1}(0)$.  Since in logarithmic terms OPE of $V'_{1,k}$ and $V_a$ is similar to the OPE of $V_{1,k}$ and $V_a$ (up to additional $q_{r,s}^{(m,n)}$ factors), it is sufficient to add the same factors for OPE of $O'_{1,k_2+1}$ and $O_{1,k_1+1}$ compared to those for OPE of $O_{1,k_2+1}$ and $O_{1,k_1+1}$:
\begin{equation}
\mathcal{O}'_{1,k_2+1}(y) \mathcal{O}_{1,k_1+1}(x) = \log |y-x|^2 \sum \limits_{s=k_2-k_1}^{k_2+k_1}  q_{0,s-k_1}^{(1,k_2+1)}(a_{1,k_1+1})\mathcal{O}_{1,1+s} + \dots
\end{equation}
where $k_2$ is in not less than $k_1$  assumed.
Thus, we reduced  the contribution to correlators with one $\mathcal{O}$-operators and three $\mathcal{W}$-s with some coefficients which can be easily calculated by performing OPEs of $\mathcal{O}$ with one of the 
$\mathcal{W}$. 
\vskip 10pt
{\bf  Contributions of  the vicinity of $y$.}
    Now we want to calculate the terms that come from the vicinity of $y$. These ones are the most tricky. 
There are two immediate problems that we see. 
First, since now we have operator $U_{k_2}(y)$, 
which is not BRST-invariant, so
$\mathcal{Q}$-exact terms in HEM  become relevant.
Second, OPE of $O'_{1,1+k_1}(x)$ with $U_{1,1+k_2}$ and, consequently, logarithmic terms in their OPE  are not as easy as in (\ref{OpWOPE}). 
However, we argue that these two problems cancel each other out in a certain sense. To see this we must take the following steps.

First, we  rewrite a product of local operators 
$U_{1,1+k_1}(x) U_{1,1+k_2}(y)$ in the path integral expectation value  
for 5-point function, using (\ref{keyrel2}), as 
 \begin{equation}
 B_{1,1+k_1}U_{1,1+k_1}(x) U_{1,1+k_2}(y)=		
   (\br{\pd}\pd  - \bar{\mathcal{Q}}\bar{B}_{-1}\pd 
  -\br{\pd}\mathcal{Q}B_{-1}+ 
\bar{\mathcal{Q}}\bar{B}_{-1}\mathcal{Q}B_{-1})
  O'_{m,n}(x)  U_{1,1+k_2}(y) \label{HEM5p}		
\end{equation}
Second, we move the action of $\mathcal{Q}$ and $\br{\mathcal{Q}}$ from $O'_{m,n}(x)$ to $U_{ 1 , 1+k_2 }(y )$ and we get in r.h.s. of (\ref{HEM5p})
\begin{equation}
\begin{aligned}
&\br{\pd}\pd O'_{m,n}(x) U_{1,1+k_2}(y) 
-\br{\pd}B_{-1} O'_{m,n}(x) \mathcal{Q}U_{1,1+k_2}(y)-\\
&{\pd} \bar{B}_{-1} O'_{m,n}(x) \bar{\mathcal{Q}}U_{1,1+k_2}(y)
+\bar{B}_{-1}B_{-1}O'_{m,n}(x) \bar{\mathcal{Q}}\mathcal{Q}U_{1,1+k_2}(y).
\end{aligned}
\end{equation}
At last, using
$\mathcal{Q}U_{1,1+k_2}(y)= \pd_y(C U_{1,1+k_2}(y))$ we obtain the following expression for the product of the fields in this piece of the 5-point correlator 
\begin{align}
&\int d^2y\,U_{k_2}(y) \int d^2x\,\pd_x \br{\pd}_x \left(H_{1,1+k_1} \br{H}_{1,1+k_1} \Theta'_{1,1+k_1} \right)\,W_{a_1}(x_1)W_{a_2}(x_2)W_{a_3}(x_3) \label{line1} \\
&-\int d^2y\, \int d^2x\,\br{\pd}_x \left(R_{1,1+k_1} \br{H}_{1,1+k_1} \Theta'_{1,1+k_1} \right) \pd_y (C U_{k_2}(y)) \,W_{a_1}(x_1)W_{a_2}(x_2)W_{a_3}(x_3) \label{line2}\\
&- \int d^2y\, \int d^2x\,\pd_x \left(\br{R}_{1,1+k_1} H_{1,1+k_1} \Theta'_{1,1+k_1} \right)\br{\pd}_y (\br{C} U_{k_2}(y))\,W_{a_1}(x_1)W_{a_2}(x_2)W_{a_3}(x_3) \label{line3}\\
&+ \int d^2y\, \int d^2x\, R_{1,1+k_1} \br{R}_{1,1+k_1} \Theta'_{1,1+k_1} \pd_y \br{\pd}_y \left(C \br{C} U_{k_2}(y)\right) \,W_{a_1}(x_1) W_{a_2}(x_2) W_{a_3}(x_3) \label{line4},
\end{align}
where $R_{1,1+k}:=B_{-1}H_{1,1+k}$.
 
All lines contain one (or both) integrals (over $x$ or $y$) that reduce to boundary contributions. Some of them correspond to the region where $x$ is close to $y$ and can be  obtained from the Stokes theorem.
These contributions are equal to the residues at the poles arising from the differentiation of logarithmic factors $\log(x-y)$, which appear in the operator expansion of the logarithmic field $V'_{1,1+k_1}(x)$ with the primary fields in $y$.
Other terms, which appear  from the differentiation 
do not contain the first-order poles and thus do not yield
 any boundary contributions at all.
\vskip 5pt

It is remarkable that in total, the boundary terms from the vicinity of $y$ is reduced to an expression similar to that obtained from the vicinity of $x_i$. 
Namely, it looks like
\begin{equation}
- \sum \limits_{s=-k_1:2}^{k_1} 
   \frac{q_{0,s}^{(1,k_1+1)}(a_{1,-k_2-1})} {\mathcal{N}(a_{1,-1-(k_2-s)})}\,
 \left\langle \int d^2y\, U_{1,(k_2-s)+1}(y)
\mathcal{W}_{a_i}(x_i) \dots \right\rangle .\label{lastcont}
\end{equation}
We will demonstrate  the derivation of  this result by a direct calculation in the Appendix for the simplest case $k_1=1$.
\vskip 5pt 
{\bf  All together.}

Bringing together (\ref{firstcont}), (\ref{continf}) and (\ref{lastcont}), we get the following expression for the (normalized) five-point correlator
\begin{align}
&C_5(a_1,a_2,a_3|k_1,k_2) = (b^{-6}-b^{-2} \left[ \Sigma_1 + \Sigma_2 + \Sigma_3 \right];  \label{5pfhemans}  \\
& \Sigma_1 =  \sum \limits_{s=-k_1:2}^{k_1} q_{0,s}^{(1,k_1 + 1)} (a_{1,-k_2-1}) \left[2 (1+k_2-s) \lambda_{1,1+k_2-s} +  \sum \limits_{i=1}^3  \sum \limits_{l=-k_2+s:2}^{k_2-s} q_{0,l}^{(1,1+k_2-s)}(a_i)\ \right]    \\
& \Sigma_2 = \sum \limits_{i=1}^3  \sum \limits_{s=-k_1:2}^{k_1} q_{0,s}^{(1,k_1+1)} (a_i) \left[2 (1+k_2) \lambda_{1,1+k_2} + \sum \limits_{l=-k_2:2}^{k_2} \left( q_{0,l}^{(1,k_2+1)}(a_i - \lambda_{0,s}) + \sum \limits_{j \neq i} q_{0,l}^{(1,k_2+1)}(a_j) \right) \right]    \\
& \Sigma_3 = 2 \lambda_{1,1+k_1} \left[ \sum \limits_{s=-k_1:2}^{k_1} \sum \limits_{l=-k_2:2}^{k_2} \left( \sum \limits_{i=1}^3  q_{0,l}^{(1,1+k_2)}(a_i) + 2 \lambda_{1,k_2+1} \right) + \sum \limits_{s=k_2-k_1}^{k_2+k_1} q_{0,s-k_1}^{(1,k_2+1)} (a_{1,k_1+1}) (1+s) \right]  \label{MLG5point}
\end{align}

{\bf Generalization to the case of $N$-point correlator} in MLG with $N>5$
can be done directly. We just need to do the same steps.
Namely, choosing of one of the $N-3$ fields of the type $U_{k_1}(y_1)$ we must  integrate over the variable $x$, using the relation (\ref{keyrel2})   for this field.

In the result we obtain boundary contributions  
of three  non-integrated fields $W_a(x_i)$, the contribution
of  $x=\infty$ and contributions of   the vicinities of $y_j$, positions of the other  $U_{k_j}(y_j), j>1$.
The final expression for the original N-point correlator will reduce to a sum of $N_ 1$-point correlators with $N_ 1<N$.


\section{Five-point correlator in  MM approach.}

Matrix models give another long-known formulation of the theory of two-dimensional gravity (for a general review see e.g. \cite{GM}. The general idea is to integrate over one or several hermitian matrices $M_i$ of size $N \times N$ with a weight $\exp (-N\, \text{Tr }V(M))$ defined by the function $V$ called "potential". There are specific values of parameters of the potential called "critical points". 
For one-matrix model they are parametrized by integer $p$. In their vicinity, a certain $N \to \infty$ limit called "double-scaling limit" can be taken. Partition function calculated in this way (as a certain function of parameters $t_k$ that define deformation of the potential away from $p$-critical point) is supposed to be connected with the generating functional in $(2,2p+1)$ MLG CFT $Z(\lambda_k) = \langle \exp \left(-\lambda_k \int d^2x\, U_k(x) \right) \rangle$, perturbative expansion of which gives the correlators studied in the first part of the work.

For some cases, agreement between results of MLG and matrix model is immediate and was known long ago \cite{MSS}. However, in the general case, to achieve coincidence one needs to perform an analytic change of coordinates from matrix model couplings $t_k$ to MLG coupling $\lambda_k$ called "resonance transformations" \cite{ABAZ}. These transformations are supposedly completely determined by requirement that derivatives of matrix model partition function $\mathcal{Z}$ with respect to $\lambda$ satisfy MLG fusion rules. For 3 and 4-point correlation numbers correspondence of matrix model answer with the one obtained from Liouville gravity was already demonstrated in \cite{ABAlZ}.
\vskip 5pt 

{\bf  Matrix-model answer for the five-point function.}

Five-point number was first calculated in \cite{GT} in Matrix model approach and is given by expressions below. 
$Z^{(I)}$ and $Z^{(J)}$ are of this form only when $\sum k_i$ is even; in the odd case they are identically zero:
\begin{align}
& Z_{k_1 k_2 k_3 k_4 k_5} = Z^{(I)} + Z^{(J)} + Z^{(1)} + Z^{(2)} \label{5pfmmans} \\
& Z^{(1)} = \sum \limits_{i=1}^5 \left(\frac{3p(p+1)}{2} F_\theta(k_i-1) -H_\theta(k_i-2) \right) - 2 \sum \limits_{i<j} F_\theta(k_i-1) F_\theta(k_j-1) - \nonumber \\
&-\frac{p(p+1)(5p^2+5p+2)}{8}; \\
& Z^{(2)} = \sum \limits_{i < j} \left(H_\theta (k_{ij}-1) - F_\theta(k_{ij}) \frac{p^2+p}{2} + F_\theta(k_{ij}) \sum \limits_{l \neq i,j} F_\theta(k_l - 1) \right) -  \sum \limits_{i<j<l} H_\theta(k_{ijl}) - \nonumber\\
&- \sum \limits_{i,j,l,m} F_\theta(k_{ij}) F_\theta(k_{lm} ); \\
& Z^{(I)} = \sum \limits_n \frac{1}{8}(2k_n - k -2) (2k_n - k -4) (2p-3-k)(2p-5-k)\, \theta (2k_n - k -6); \\
& Z^{(J)} = \sum \limits_{i<j} \left(H(k-k_{ij}) - \frac{(k-2k_{ij})(k-2k_{ij}+2)(2p-3-k)(2p-5-k)}{8} \theta(2k_{ij}-k-2) \right) \times \\ \nonumber
& \times \theta (p-1-k+k_{ij}) \theta(p-1-k_{ij})
\end{align}
Here we have introduced notations
\begin{align}
& F_\theta(k) = \frac{1}{2} (p-k-1)(p-k-2) \theta(p-2-k),\,H_\theta(k) = \frac{1}{2} F_\theta(k) F_\theta(k+2); \\
& k = \sum \limits_{i=1}^5 k_i,\,k_{ij} = k_i + k_j,\,k_{lmn} = k_l + k_m + k_n
\end{align}
($H$ is the same as $H_\theta$ without the theta-function). In general case this expression is quite complicated, but it can be somewhat simplified in the region of parameter space when for any $i \neq j \neq l$ $k_{ijl}<p$; it factorizes to be
\begin{align}
&\frac{Z_{k_1 k_2 k_3 k_4 k_5}}{(2p-3-k)(2p-5-k)}=\frac{1}{8} \left(4 \sum_i k_i^2 - k^2 -2k-8 - \right. \nonumber \\
&\left. \sum \limits_{m<n}\theta(2k_{mn} - k - 2) (k-2k_{mn})(k-2k_{mn}+2) + \sum \limits_n (2k_n - k -2) (2k_n - k -4) \right) \label{FLanswer}
\end{align}
Interesting feature of this answer is the factor in the r.h.s. which is actually equal to the number of conformal blocks in the minimal model part of the corresponding MLG correlator \cite{artemev2022}. 
When this number is maximal and equal to $(1+k_1)(1+k_2)$, this answer agrees with the expression obtained by Fateev and Litvinov in \cite{fatlit2008} from CFT calculation using Coulomb integrals.
\vskip 5pt

{\bf  Comparison with matrix model approach (some examples).}

To perform comparison with (\ref{5pfmmans}), we need to substitute in (\ref{5pfhemans}) $a_i$ corresponding to dressed degenerate fields $a_{1,-k_i-1}$ and also $b = \sqrt{2/(2p+1)}$. We suppose that parameters are ordered as $0 \leq k_1 \leq k_2 \leq k_3 \leq k_4 \leq k_5 \leq p-1$. We will denote the  correlator normalized as in (\ref{5pfhemans}) evaluated at such values of $a$ as $b^2 \times Z^{(HEM)}_{k_1,k_2,k_3,k_4,k_5}$.

 As it was the case for the four-point function, we expect the  coincidence only for the cases when number of conformal blocks in the considered five-point correlator is specific; analogy with the four-point correlator suggests that it should be maximal and equal to $(1+k_1)(1+k_2)$. 

However, the MLG and MM expressions generally do not match even under this condition. For example, for the case when
$({k_1,k_2,k_3,k_4,k_5})=({1,1,2,k-1,k-1}$) and large enough $p$ (more precisely, $p>2k+3$) in MLG the expression for 5-point function (\ref{5pfhemans}) looks like
\begin{equation}
Z_{MLG}=8p^2- 16pk+8k^2-48p+50k+85,
\end{equation}
while in MM model it is
\begin{equation}
Z_{MM}=8p^2- 16pk+8k^2-48p+48k+70.
\end{equation}

 


\section{Acknowledgements.}
The work was carried out at  Landau  Institute of Theoretical Physics in the framework of the state assignment No.  0029-2019-0004 

\section{Appendix}

{\bf Check of the result in 5-point correlator case for $k_1=1$.}

We will illustrate the statement in section 3 for the simplest case $k_1 = 1$. We will need the explicit form of $H_{1,2}$, the fact that $R_{1,2} = b^2 B$ and OPEs
\begin{align}
&V'_{1,2}(x) V_{a}(y) = \log(|x-y|^2) \left[ \overbrace{|x-y|^{2ab} \tilde{C}_L^+(a) [V_{a-b/2}(y)]}^{(1)} + \overbrace{|x-y|^{2(1 -ab + b^2)} \tilde{C}_L^-(a) [V_{a+b/2}(y)]}^{(2)} \right]  \\
&\Phi_{1,2}(x) \Phi_{a-b} (y) =  \underbrace{|x-y|^{2(ab-b^2)} C_M^+ (a-b)[\Phi_{a-b/2}(y)]}_{(3)} + \underbrace{ |x-y|^{2(1-ab)} C_M^-(a-b) [\Phi_{a-3b/2}(y)]}_{(4)} \label{fiope}
\end{align}
We denote $a = b+ \alpha_{1,k_2+1}$ and $\tilde{C}_L$ are Liouville structure constants decorated by $q$-factors as in (\ref{vprime12ope}). Although in our case $\Phi_{a-b} = \Phi_{1,k_2+1}$, we assume that $k_2>k_1$ (as before) so that the second OPE looks the same as in non-degenerate case. Multiplying these two OPEs and acting by $H_{1,2} \br{H}_{1,2}$, we get the OPE of $O'_{1,1+k_1}$ and $U_{k_2}$; on the level of primary fields and when considering terms where derivatives in $H$ do not act on the logarithm (we need to leave it intact so that the second derivative acts on it and gives the delta-function), the OPE is given by
\begin{align}
\log |x-y|^2 &\left[\tilde{C}^+_L C^+_M \left(-\frac{b^2}{x-y} + b^2 CB\right) \left(-\frac{b^2}{\br{x-y}} + b^2 \br{CB}\right) |x-y|^{2(2ab-b^2)} V_{a-b/2} \Phi_{a-b/2} (y) +  \right. \nonumber\\
&\left.\tilde{C}^+_L C^-_M \left(\frac{1-2ab}{x-y} + b^2 CB\right) \left(\frac{1-2ab}{\br{x-y}} + b^2 \br{CB}\right) |x-y|^{2} U_{a-b/2} (y) +  \right. \nonumber \\
&\left.\tilde{C}^-_L C^+_M \left(\frac{2ab - 2 b^2 - 1}{x-y} + b^2 CB\right) \left(\frac{2a b - 2 b^2 - 1}{\br{x-y}} + b^2 \br{CB}\right) |x-y|^{2} U_{a+b/2} (y) + \right. \nonumber \\
&\left.\tilde{C}^-_L C^-_M \left(-\frac{b^2}{x-y} + b^2 CB\right) \left(-\frac{b^2}{\br{x-y}} + b^2 \br{CB}\right) |x-y|^{2(2-2ab+b^2)} V_{a+b/2} \Phi_{a-3b/2} (y) \right] 
\end{align}
We take the derivative $\pd_x \br{\pd}_x$ of this to calculate one of boundary contributions to (\ref{line1}). Due to absence of BRST-ghosts $C$ terms not of the form $U_\# = V_\# \Phi_{\#-b}$ in the OPE do not cancel and, moreover, there are ghosts in the "correct" terms. 
However, the multipliers for additional terms are strongly 
reminiscent of the OPE
\begin{equation}
R_{1,2}(x) C(y) = \frac{b^2}{x-y} + b^2 BC + \dots = -\left(-\frac{b^2}{x-y} + b^2 CB \right) + \dots
\end{equation}
This is what would appear if we consider terms in lines (\ref{line2}), (\ref{line3}), (\ref{line4}).  E.g. the contribution of (\ref{line3}) is given by
\begin{align*}
-\pd_x \br{\pd}_y \log |x-y|^2 &\left[ \tilde{C}^+_L C^+_M\left(\frac{b^2}{x-y} - b^2 CB\right) \left(-\frac{b^2}{\br{x-y}} + b^2 \br{CB}\right) |x-y|^{2(2ab-b^2)} V_{a-b/2} \Phi_{a-b/2} (y) +  \right.\\
&\left. \tilde{C}^+_L C^-_M\left(\frac{b^2}{x-y} - b^2 CB\right) \left(\frac{1-2ab}{\br{x-y}} + b^2 \br{CB}\right) |x-y|^{2} U_{a-b/2} (y) +  \right.\\
&\left.\tilde{C}_-^L C_+^M \left(\frac{b^2}{x-y} - b^2 CB\right) \left(\frac{2a b - 2 b^2 - 1}{\br{x-y}} + b^2 \br{CB}\right) |x-y|^{2} U_{a+b/2} (y) + \right.\\
&\left.\tilde{C}^-_L C^-_M \left(\frac{b^2}{x-y} - b^2 CB\right) \left(-\frac{b^2}{\br{x-y}} + b^2 \br{CB}\right) |x-y|^{2(2-2ab+b^2)} V_{a+b/2} \Phi_{a-3b/2} (y) \right]
\end{align*}
On the level of primary fields where we only differentiate the OPE coefficients (we work only on this level anyway) we can rewrite $\br{\pd}_y$ as $-\br{\pd}_x$; after that contributions proportional to $V_{a-b/2} \Phi_{a-b/2}$ and $V_{a+b/2} \Phi_{a-\frac{3b}{2}}$ cancel inside of the derivative. Same cancellation happens for these terms coming from (\ref{line3}) and (\ref{line4}) that we will not write for brevity. On the other hand, if we sum contribution from all 4 lines for e.g. factor before $U_{a-b/2}$ we get
\begin{align}
& \left(\frac{1-2ab}{x-y} + b^2 CB\right) \left(\frac{1-2ab}{\br{x-y}} + b^2 \br{CB}\right) + \left(\frac{1-2ab}{x-y} + b^2 CB\right) \left(\frac{b^2}{\br{x-y}} - b^2 \br{CB}\right) + \nonumber \\
&+ \left(\frac{b^2}{x-y} - b^2 CB\right) \left(\frac{1-2ab}{\br{x-y}} + b^2 \br{CB}\right) + \left(\frac{b^2}{x-y} - b^2 CB\right) \left(\frac{b^2}{\br{x-y}} - b^2 \br{CB}\right) =  \nonumber \\
&= \left(\frac{1-2ab}{x-y} +  b^2 CB + \frac{b^2}{x-y} - b^2 CB \right) \left(\frac{1-2ab}{\br{x-y}} +  b^2 \br{CB} + \frac{b^2}{\br{x-y}} - b^2 \br{CB} \right) =  \nonumber \\
&= \frac{1}{|x-y|^2} (1-2ab + b^2)^2
\end{align}
The ghosts cancel and we obtain the same factor that we would get for the OPE with $W$.
\vskip 5pt


\end{document}